\documentclass[pre,aps,showpacs,preprint]{revtex4}
\usepackage{epsfig}
\usepackage[T1]{fontenc}
\usepackage{ae,aecompl}

\usepackage{amsfonts}
\usepackage{amsmath}
\usepackage{amssymb}
\usepackage{graphicx}%

\usepackage{indentfirst}


\begin{document}

\title{Three-dimensional multi-mesh material point method for solving collision problems\footnotetext{Supported by National Natural
Science Foundation of China (contract number: 10472052) and Science
Foundation of National Key Laboratory of Computational Physics,
Institute of Applied Physics and Computational Mathematics, Beijing,
China. }}

\author{PAN Xiaofei, XU Aiguo, ZHANG Guangcai and ZHU Jianshi}
\affiliation{Laboratory of Computational Physics, \\Institute of
Applied Physics and Computational Mathematics, P. O. Box 8009-26,
Beijing 100088, P. R. China}
\author{MA Shang and ZHANG Xiong}
\affiliation{School of Aerospace, Tsinghua University, Beijing
100084, P. R. China}
\date{\today}

\noindent

\begin{abstract}
Contact algorithm between different bodies plays an important role
in
  solving collision problems. Usually it is not easy to be treated very well.
  Several ones for material point method
  were proposed by Bardenhangen, Brackbill, and
  Sulsky\cite{Bardenhagen2000,Bardenhagen2001}, Hu and Chen\cite{Hu_Chen2003}.
  An improved one for three-dimensional material point method is presented in this paper.
  The improved algorithm emphasizes the energy conservation of the system and faithfully recovers opposite acting forces
  between contacting bodies. Contrasted to the one by Bardenhagen, both the normal
  and tangential contacting forces are more appropriately applied to
  the contacting bodies via the contacting nodes of the background
  mesh; Contrasted to the one by Hu and Chen, not only the
  tangential velocities but also the normal ones are handled
  separately in respective individual mesh. This treatment ensures
  not only the contact/sliding/separation procedure but also the friction between contacting bodies are recovered.
  The presented contact algorithm is validated via numerical
  experiments including rolling simulation, impact of elastic spheres, impact of a Taylor bar and
  impact of plastic spheres. The numerical results show that the multi-mesh material point method with the
  improved contact algorithm is more suitable for solving collision problems.
\end{abstract}


\maketitle

\section{INTRODUCTION}

Phenomena with large deformation and/or large rotation are very common in
nature, especially in fields of hypervelocity impact and explosion.
Numerical simulations of such processes are necessary and challenging. The
material point method(MPM) is an extension of FLIP\cite{Brackbill86,
Brackbill88} which combines the strength of Eulerian and Lagrangian
descriptions of the material, to the solid mechanics. The Lagrangian
description is provided by discretizing each body by a collection of
material points, and the Eulerian description is based on a background
computational mesh. Information carried by the material points is projected
on to the background mesh where equations of motion are solved. The mesh
solution is then used to update the material points. In Sulsky et al\cite%
{Sulsky94,Sulsky95} a weak formulation of the MPM algorithm for solid
mechanics is given and the method is framed in the terms of finite elements.
The MPM combines the advantages of Eulerian and Lagrangian methods, which
can avoid the distortion of Lagrangian mesh and track the boundaries of
bodies. The method has been applied to the large strain problems\cite%
{Wiechowski2004,Coetzee2005}, calculations with dynamical energy release rate%
\cite{Tan2002}, fracture mechanics\cite{Joris2005}, dynamics failure\cite%
{ZChen2002, ZChen2003}, hypervelocity impact\cite{xzhang2006}, the thin
membranes\cite{Allen1999}, granular materials\cite%
{Bardenhagen2000,Cummins2002,Bardenhagen1998,Bardenhagen2000-2} etc.

In MPM, no slip contact between bodies is contained in the basic algorithm
without additional cost. But at most cases separation or sliding may happen
during the moving of bodies. A contact algorithm was presented by
Bardenhagen, Brackbill and Sulsky to simulate the interactions of the grains
of granular material\cite{Bardenhagen2000}. In the algorithm, the contact
may occur if the material points of different bodies are projected on to the
same nodes of the background mesh, and the contact force is associated with
the center-of-mass velocity. Bardenhagen's algorithm is linear in the number
of grains and allows separation, sliding and rolling. With the contact
algorithm MPM has been successful in simulating the large deformation of
shear in granular material, having an advantage over traditional finite
element methods(FEM) in that the use of regular grid eliminates the
necessity to do costly search for contact surfaces. In order to apply MPM to
stress propagation in the granular material, the contact algorithm is
improved by Bardenhagen et al\cite{Bardenhagen2001}. In Bardenhagen's
improved contact algorithm, the normal traction is included in the contact
logic to more appropriately determine the free separation criterion.

To release the no-slip contact algorithm in MPM, a multi-mesh mapping scheme
is proposed by Hu and Chen\cite{Hu_Chen2003}. In the multi-mesh mapping
scheme, each material lies in an individual background mesh rather than in
the common background mesh. The meshing process of spur gears is simulated
by Hu and Chen with their contact algorithm. To avoid interpenetration and
allow separation in the gear meshing process, the normal velocity of any
particle at the contact surface is calculated in the common background mesh,
while the tangential velocity is found based on the corresponding
information in respective individual mesh. With the proposed contact
algorithm, Hu and Chen have successfully simulated the contact and
separation of the gears. In their scheme, normal acceleration is set to be
equal if particles of different bodies are mapped on the same node. But for
some cases, the bodies may separate although their particles are still
mapped on the same nodes, which can cause energy dissipation during contact.
In the contact algorithm by Hu and Chen, the friction between different
bodies is completely ignored because the tangential velocities of different
bodies are assumed to be independent.

In this paper, an improved multi-mesh contact algorithm for
three-dimensional MPM is proposed. In the present contact algorithm, the
criterion of contact condition is similar to Bardengen's which ensures that
the search for the contact of different bodies is fast, but the multi-mesh
is used to calculate the normal and tangential velocities of different
bodies. To avoid interpenetration the normal contact force is calculated at
the contact surface and the Coulomb friction is applied in the tangential
direction. Contrasting with Bardenhagen's algorithm, both the normal and
tangential contacting force are more appropriately applied to the contacting
bodies at the contacting nodes of the background mesh; Contrasting with Hu
and Chen's contact algorithm, the normal velocities of different bodies are
deal with separately just as the tangential velocities to not only ensure
the contact/sliding/separation procedure can be simulated but also ensure
the friction between different bodies can be applied. With the presented
algorithm the total energy of the system is nearly constant during both
elastic and non-elastic collision procedures, which shows numerical energy
dissipation is little.

This paper is organized as follows. The material point method is
briefly reviewed in section \ref{SEC-MPM}, and the
new multi-mesh contact algorithm is illuminated in Section \ref{SEC-CONTACT}%
. Several numerical examples are presented In section \ref{SEC-NUMERICAL} to
validate the interaction between bodies with the contact algorithm.
Numerical results obtained by the proposed contact algorithm presented are
compared with those obtained by Bardenhagen's contact algorithm which show
the proposed algorithm is more suitable in solving collision problems in
which the numerical energy dissipation need to be very low. Section \ref%
{SEC-CON} concludes the paper with some remarks and observations.

\section{The Material Point Method\label{SEC-MPM}}

The MPM is a particle method based on particle-in-cell(PIC) method in
computational fluid mechanics. The method was initially developed for and
has been successfully applied in problems involving large-deformation, large
rotations of solid, etc. For continuum bodies, the conservation equation for
mass is
\begin{equation}
\frac{\mathrm{d}\rho}{\mathrm{d}t} + \rho \nabla \cdot \mathbf{v} = 0.
\label{eq-pmp-mass}
\end{equation}
And for pure mechanical problems the differential equation of balance is
\begin{equation}
\rho \frac{\mathrm{d}\mathbf{v} }{\mathrm{d}t} = \nabla \cdot \mathbf{\sigma}
+ \rho \mathbf{b},  \label{eq-pmp-moment}
\end{equation}
where $\rho$ is the mass density, $\mathbf{v}$ is the velocity, $\mathbf{%
\sigma}$ is the stress tensor and $\mathbf{b}$ is the body force.

The formulation \eqref{eq-pmp-moment} is solved in a Lagrangian frame on a
finite element mesh. The Lagrangian formulation means that the momentum
equation does not contain the convection term which can cause significant
numerical error in pure Eulerian approaches.

In MPM, the continuum bodies are discretized with $N_p$ material particles.
Each material particle carries the information of position, velocity, mass,
density, stress, strain and all other internal state variables necessary for
the constitutive model. Since the mass of each material particle is equal
and fixed, Eq. \eqref{eq-pmp-mass} is automatically satisfied. At each time
step, the mass and velocities of the material particles are mapped onto the
background computational mesh(grid). The mapped nodal velocity $\mathbf{v}_j$
is obtained through the following equation,
\begin{equation}
\sum_{j}{m_{ij}\mathbf{v}_j} = \sum_{p}{m_p \mathbf{v}_p N_{i}(\mathbf{x}_p)}
\label{eq-pmp-velo}
\end{equation}
where $m_p$, $\mathbf{v}_p$ and $\mathbf{x}_p$ are the mass, velocity and
position of particle $p$, separately. $N_i$ is the element shape function, $i
$ and $j$ indexes of node. For three-dimensional problem, a eight-node cell
is employed with the shape functions given by
\begin{equation}
N_i = \frac{1}{8}(1+\xi\xi_{i})(1+\eta\eta_{i})(1+\zeta\zeta_{i})
\label{eq-pmp-shape}
\end{equation}
where $\xi$, $\eta$ and $\zeta$ are the natural coordinates of a material
particle in the cell along the x-, y- and z-directions, $\xi_{i}$, $\eta_{i}$
and $\zeta_{i}$ are the natural coordinates of the node $i$ in the cell
along the three directions.

In the Eq. \eqref{eq-pmp-velo}, the consistent mass matrix, $m_{ij}$, is
\begin{equation}
m_{ij} = \sum_{p} m_p N_{i}(\mathbf{x}_p) N_{j}(\mathbf{x}_p)
\label{eq-pmp-mass-matrix}
\end{equation}
In practice, we generally replace $m_{ij}$ with a lumped, diagonal mass
matrix so that Eq. \eqref{eq-pmp-velo} becomes
\begin{equation}
m_{i}\mathbf{v}_i = \sum_{p} m_p \mathbf{v}_p N_i( \mathbf{x}_p )
\label{eq-pmp-mass-velo-2}
\end{equation}
where lumped mass is
\begin{equation}
m_{i} = \sum_p m_p N_i(\mathbf{x}_p)  \label{eq-pmp-mass-2}
\end{equation}

After the information is mapped from material particles to mesh nodes, the
discrete formulation of Eq. \eqref{eq-pmp-moment} on the mesh nodes can be
obtained, as described below.

The weak form of Eq. \eqref{eq-pmp-moment} can be found, based on the
standard procedure used in the finite element method,

\begin{equation}
 \int_{\Omega }{ \rho \delta \mathbf{v}\cdot \mathrm{d}\mathbf{v/}\mathrm{d}t%
\mathrm{d}\Omega }+ \int_{\Omega }{\delta (\mathbf{v}\nabla) \cdot
\boldsymbol{\sigma}\mathrm{d}\Omega } - \int_{\Gamma _{t}}{\ \delta
\mathbf{v}\cdot \mathbf{t}\mathrm{d}\Gamma }
  -\int_{\Omega
}{\ \rho \delta \mathbf{v}\cdot \mathbf{b}\mathrm{d}\Omega } =0
\texttt{.} \label{1}
\end{equation}
where $\Omega $ is the domain to be solved, $\Gamma _{t}$ is the traction
boundary, $\mathbf{\sigma }$ is the stress tensor, $\mathbf{t}$ is the
external force vector and $\mathbf{b}$ is the body force vector.

Since the continuum bodies is described with the use of a finite set of
material particles, the mass density can be written as,
\begin{equation}
\rho(\mathbf{x}) = \sum_{p=1}^{N_p}{\ m_p\delta(\mathbf{x}-\mathbf{x}_p) }
\label{eq-pmp-density}
\end{equation}
where $\delta$ is the Dirac delta function with dimension of the inverse of
volume. The substitution of Eq. \eqref{eq-pmp-density} into Eq. %
\eqref{eq-pmp-weakform} converts the integral to the sums of quantities
evaluated at the material particles, namely,

\begin{equation}
m_{i} \frac{\mathrm{d}\mathbf{v}_i}{\mathrm{d}t} = (\mathbf{f}_i)^{\mathrm{%
int}} + (\mathbf{f}_i)^{\mathrm{ext}}  \label{eq-pmp-solve}
\end{equation}
where $m_i$ is the lumped mass, $(\mathbf{f}_i)^{\mathrm{int}}$ and $(%
\mathbf{f}_i)^{\mathrm{ext}}$ are the external force vector and internal
force vector which read separately

\begin{align}
(\mathbf{f}_i)^{\mathrm{int}} & = - \sum_{p}^{N_p} {\ m_p \mathbf{\sigma}
\cdot (\nabla N_i) / \rho_p },  \label{eq-pmp-fint} \\
(\mathbf{f}_i)^{\mathrm{ext}} & = \sum_{p=1}^{N_p}{\ N_{i}\mathbf{b}_p +
\mathbf{f}_i^c }  \label{eq-pmp-fext}
\end{align}
where the vector $\mathbf{f}_i^c$ is the contact force which is the external
nodal force not including the body force and is illustrated in the following
section.

An explicit time integrator is used to solve Eq. \eqref{eq-pmp-solve} for
the nodal accelerations, with the time step satisfying the stability
condition. The critical time step is the ratio of the smallest cell size to
the wave speed. After the equations of motion are solved on the cell nodes,
the new nodal values of acceleration are used to update the velocity of the
material particles. The strain increment for each material particle is
determined with the use of gradient of the nodal basis function evaluated at
the material particle position. The corresponding stress increment can be
found from the constitutive model. The internal state variables can also be
completely updated. The computational mesh may be discarded, and a new mesh
is defined, if desired for the next time step. As a result, an effective
computational mesh could be chosen for convenience.

\section{The contact algorithm\label{SEC-CONTACT}}

The MPM with a natural no-slip contact algorithm is based on a common
background mesh. As a result, it is impossible to separate the contacting
bodies. Bardenhagen et al.\cite{Bardenhagen2000,Bardenhagen2001} have
proposed a contact algorithm in which the contact between bodies is handled
when the velocity field of an individual particle differs from the single,
center-of-mass velocity field in the cell containing contacting particles.
Their contact algorithm was incorporated into the MPM to simulate the
interactions in granular materials based on the velocity field.

In this section, we will improve the multi-mesh contact algorithm which
recovers more faithfully the opposite acting forces between contacting
bodies. In MPM, several bodies may be mapped on to the same nodes of the
background mesh, so it is necessary to define a multi-value of velocity and
mass on every node. In practice, it is impossible that the number of values
defined at one node is as many as the number of bodies, otherwise the memory
of computer will be too much wasted if there are thousands of bodies to be
simulated. In this paper we define four values on every node. That is to
say, there are at most four bodies mapped on to the same nodes, although
there can be thousands of bodies in the whole domain. In that case, each
node has a mesh mass $m_{i}^{g}$ and momentum $\mathbf{P}_{i}^{g}$
associated with it, where $g$ ranges from one to four and the mesh velocity $%
\mathbf{v}_{i}^{g}$ can be obtained from the mesh momentum and the mass,
\begin{equation}
\mathbf{v}_{i}^{g}=\mathbf{P}_{i}^{g}/m_{i}^{g}  \label{eq-contact-rate-0}
\end{equation}%
Note that if the mesh mass $m_{i}^{g}$ is close to zero, the obtained mesh
velocity maybe singular which will cause error during the calculations. In
this paper, to avoid the singularity, the shape function is altered, if $\xi
$, $\eta $ or $\zeta $ is small than $-0.99$ or larger than $0.99$ that
means the material point is too close to the node, $\xi $, $\eta $ or $\zeta
$ is adjusted to $-0.99$ or $0.99$. Since the shape functions have compact
support, only those nodes in the vicinity of the bodies will have a
meaningful velocity and the body velocity at other nodes will be zero.

Obviously, if the momenta of two bodies are projected on to the same node,
the contact may occur and the contact between bodies $r$ and $s$ is directed
by comparing the fields $\mathbf{v}_i^r$ and $\mathbf{v}_i^s$ which are
determined by using mass weighting given in Eq. \eqref{eq-contact-rate-0},
\begin{equation}
(\mathbf{v}_i^r-\mathbf{v}_i^s)\cdot \mathbf{n}_i^{rs} > 0,
\label{eq-contact-rate-1}
\end{equation}
where $\mathbf{n}_i^{rs}$ is the unit outward normal at node $i$ along the
boundary. Multiply Eq. \eqref{eq-contact-rate-1} with $m_i^r m_i^s$, it can
be written as,
\begin{equation}
(m_i^s \mathbf{P}_i^r - m_i^r\mathbf{P}_i^s)\cdot \mathbf{n}_i^{rs} > 0
\label{eq-contact-rate-1a}
\end{equation}
If Eq. \eqref{eq-contact-rate-1} is satisfied, the velocities of body $r$
and body $s$ are adjusted to new values $\bar{\mathbf{v}_i^r}$ and $\bar{%
\mathbf{v}_i^s}$ so that
\begin{equation}
\bar{\mathbf{v}}_i^r \cdot \mathbf{n}_i^{rs} = \bar{\mathbf{v}}_i^s\cdot
\mathbf{n}_i^{rs}  \label{eq-contact-rate-2}
\end{equation}
holds. That is, the normal components of velocity of body $r$ and body $s$
are set to be equal. Eq. \eqref{eq-contact-rate-2} can also be written as
\begin{equation}
m_i^s \bar{\mathbf{P}}_i^r \cdot \mathbf{n}_i^{rs} = m_i^r \bar{\mathbf{P}}%
_i^{s}\cdot \mathbf{n}_i^{rs}.  \label{eq-contact-rate-2a}
\end{equation}
As a reasonable contact algorithm, the momentum is required to be unaltered,
i.e.,
\begin{equation}
(\mathbf{P}_i^r + \mathbf{P}_i^s) \cdot\mathbf{n}_i^{rs} = (\bar{\mathbf{P}}%
_i^r + \bar{\mathbf{P}}_i^s)\cdot \mathbf{n}_i^{rs}
\label{eq-contact-rate-3}
\end{equation}
From Eqs. \eqref{eq-contact-rate-2a} and \eqref{eq-contact-rate-3} the
updated mesh momenta are obtained,
\begin{equation}
\bar{\mathbf{P}}_i^r = \mathbf{P}_i^r - ( m_i^s\mathbf{P}_i^r - m_i^r\mathbf{%
P}_i^s )\cdot \mathbf{n}_i^{rs}\mathbf{n}_i^{rs}/( m_i^r + m_i^s ),
\label{eq-contact-rate-4a}
\end{equation}
\begin{equation}
\bar{\mathbf{P}}_i^s = \mathbf{P}_i^s + ( m_i^s\mathbf{P}_i^r - m_i^r\mathbf{%
P}_i^s )\cdot \mathbf{n}_i^{rs}\mathbf{n}_i^{rs}/( m_i^r + m_i^s ).
\label{eq-contact-rate-4b}
\end{equation}
So the updated mesh velocities are:
\begin{equation}
\bar{\mathbf{v}}_i^r = \mathbf{v}_i^r - m_i^s(\mathbf{v}_i^r - \mathbf{v}%
_i^s )\cdot \mathbf{n}_i^{rs}\mathbf{n}_i^{rs}/( m_i^r + m_i^s ),
\label{eq-contact-rate-4c}
\end{equation}
\begin{equation}
\bar{\mathbf{v}}_i^s = \mathbf{v}_i^s + m_i^r( \mathbf{v}_i^r - \mathbf{v}%
_i^s )\cdot \mathbf{n}_i^{rs}\mathbf{n}_i^{rs}/( m_i^r + m_i^s ).
\label{eq-contact-rate-4d}
\end{equation}
Especially, if body $s$ is a rigid wall, we set the value of $m_i^s$ much
larger than that of $m_i^r$. Thus Eq. \eqref{eq-contact-rate-4c} and Eq. %
\eqref{eq-contact-rate-4d} can be reduced to:
\begin{equation}
\bar{\mathbf{v}}_i^r = \mathbf{v}_i^r - (\mathbf{v}_i^r - \mathbf{v}_i^s
)\cdot \mathbf{n}_i^{rs},  \label{eq-contact-rate-4e}
\end{equation}
\begin{equation}
\bar{\mathbf{v}}_i^s = \mathbf{v}_i^s ,  \label{eq-contact-rate-4f}
\end{equation}
Obviously, the velocity of rigid body $s$ is not altered during the contact.

The equations \eqref{eq-contact-rate-4c} and \eqref{eq-contact-rate-4d}
determining the velocities are identical to that by Bardenhagen\cite%
{Bardenhagen2000} and in practice they are same, but the calculation of the
normal and tangential contact forces makes the difference which will be
described as following are different.

Once bodies $r$ and $s$ contact, they move together along the normal until
they separate when the contact condition expressed in Eq. %
\eqref{eq-contact-rate-1a} is not satisfied. So the acceleration along the
normal of body $r$ is equal to that of $s$ during the course of the contact.
That is
\begin{equation}
\mathbf{a}_i^r \cdot \mathbf{n}_i^{rs} = \mathbf{a}_i^s \cdot \mathbf{n}%
_i^{rs}  \label{eq-contact-rate-5}
\end{equation}
where $\mathbf{a}_i^r$ and $\mathbf{a}_i^s$ are the accelerations of bodies $%
r$ and $s$ at node $i$, respectively. They can be obtained from the
Newtonian second law,
\begin{equation}
m_i^r (\mathbf{a}_i^r \cdot \mathbf{n}_i^{rs}) = \mathbf{f}_i^{r,\mathrm{int}%
} \cdot \mathbf{n}_i^{rs} - f_i^{\mathrm{nor}}  \label{eq-contact-rate-6a}
\end{equation}
\begin{equation}
m_i^s (\mathbf{a}_i^s \cdot \mathbf{n}_i^{rs}) = \mathbf{f}_i^{s,\mathrm{int}%
} \cdot \mathbf{n}_i^{rs} + f_i^{\mathrm{nor}}  \label{eq-contact-rate-6b}
\end{equation}
where $f_i^{nor}$ is the normal contact force between body $r$ and body $s$
at node $i$, which can be obtained from Eq. \eqref{eq-contact-rate-5}--%
\eqref{eq-contact-rate-6b},
\begin{equation}
f_i^{\mathrm{nor}} = \big(m_i^s \mathbf{f}_i^{r,{\mathrm{int}}} - m_i^r
\mathbf{f}_i^{s,{\mathrm{int}}}\big) \cdot \mathbf{n}_i^{rs}/(m_i^r + m_i^s)
\label{eq-contact-rate-7}
\end{equation}

Note that the normal contact force must be nonnegative. So once $f_i^{%
\mathrm{nor}}$ is negative, it is set to be zero. That is
\begin{equation}
f_i^{\mathrm{nor}} = \left\{
\begin{array}{l}
\Psi /(m_i^r + m_i^s), \hspace{0.5cm} \Psi \ge 0 \\
0, \hspace{2.4cm} \Psi < 0%
\end{array}%
. \right.  \label{eq-contact-rate-8}
\end{equation}
where $\Psi = \big(m_i^s \mathbf{f}_i^{r,{\mathrm{int}}} - m_i^r \mathbf{f}%
_i^{s,{\mathrm{int}}}\big) \cdot \mathbf{n}_i^{rs}$. For the cases $f_i^{%
\mathrm{nor}}$ in Eq. \eqref{eq-contact-rate-7} is positive, the contacting
bodies at time $t$ may still contact in the next time step although the
criterion of contact in Eq.\eqref{eq-contact-rate-1a} is not satisfied, so
the contact condition should be applied in the next time step.

The contact force in the Bardenhagen's contact algorithm is calculated as
\begin{equation}
f_i^{r, \mathrm{nor}} = m_i^{r} \left[ (\widetilde{\mathbf{v}}_i - {\mathbf{v%
}}_i^{r}) \cdot \mathbf{n}_i^{rs}\right] / \Delta t
\label{eq-contact-rate-8a}
\end{equation}
where $\widetilde{\mathbf{v}}_i$ is the center-of-mass velocity at node $i$.
Actually, the normal contact force may still be very large even if the
normal velocities of different bodies at contact nodes are same during the
course of contacting. The normal contact force calculated by %
\eqref{eq-contact-rate-8a} is not physical and may cause numerical energy
dissipation which will be shown in the next section.

When without friction, the contact algorithm has been finished up to now. In
the case with friction, the frictional slip is accomplished by adjusting the
tangential component. To apply Coulomb friction, we first calculate the
force necessary to cause the bodies to stick together completely. Again, the
comparison of the mesh velocity of body $r$ to that of body $s$ provides
exactly the correct constraint for no-slip contact if body $r$ and body $s$
contact, the relative tangential velocity is
\begin{equation}
(\mathbf{v}_i^r - \mathbf{v}_i^s) \cdot \mathbf{s}_i^r
\label{eq-contact-rate-9}
\end{equation}
where $\mathbf{s}_i^r$ is the unit tangential at node $i$ along the
boundary,
\begin{equation}
\mathbf{s}_i^r = \big(( \mathbf{v}_i^r - \mathbf{v}_i^s ) - ( \mathbf{v}_i^r
- \mathbf{v}_i^s ) \cdot \mathbf{n}_i^{rs} \mathbf{n}_i^{rs} \big ) / |\big(%
( \mathbf{v}_i^r - \mathbf{v}_i^s ) - ( \mathbf{v}_i^r - \mathbf{v}_i^s )
\cdot \mathbf{n}_i^{rs} \mathbf{n}_i^{rs} \big )|  \label{eq-contact-rate-10}
\end{equation}

To get an appropriate frictional force allowing slip, we start from the
non-slip condition. The tangential velocities of body $r$ and body $s$ must
be set to be equal after one time step $\Delta t$. Suppose the tangential
contact force is $f_i^{\mathrm{tang}}$. It should satisfy
\begin{equation}
\big(\mathbf{P}_i^r \cdot\mathbf{s}_i^r + (\mathbf{f}_i^{r,{\mathrm{int}}%
}\cdot{\mathbf{s}}_i^r - f_i^{\mathrm{tang}}) \Delta t \big) / m_i^r = \big(%
\mathbf{P}_i^s \cdot\mathbf{s}_i^r + (\mathbf{f}_i^{s,{\mathrm{int}}}\cdot%
\mathbf{s}_i^r + f_i^{\mathrm{tang}}) \Delta t \big) / m_i^s.
\label{eq-contact-rate-11}
\end{equation}
Then the needed constraining tangential force $f_i^{\mathrm{tang}}$ is
\begin{equation}
f_i^{\mathrm{tang}} = \big(( m_i^s \mathbf{P}_i^r - m_i^r \mathbf{P}_i^s ) +
( m_i^s\mathbf{f}_i^{r,{\mathrm{int}}} - m_i^r\mathbf{f}_i^{s,{\mathrm{int}}%
} ) \Delta t \big)\cdot \mathbf{s}_{i}^{r} / \big( (m_i^r + m_i^s)\Delta t %
\big)  \label{eq-contact-rate-12}
\end{equation}
The expected frictional force should equal to $f_i^{\mathrm{tang}}$ if the
latter is small, and be proportional to the magnitude of the normal force
and independent of the contact area if $f_i^{\mathrm{tang}}$ exceed a
specified value. That is to say, the frictional force just balances the
tangential force to prevent relative tangential motion when the latter is
small. When the latter is large, we limit the frictional force to have a
magnitude less than it to allow tangential slip between the contacting
bodies. The direction of the frictional force is chosen as in %
\eqref{eq-contact-rate-10} to oppose the relative motion. Putting these
requirements together yields,
\begin{equation}
f_i^{\mathrm{fric}} = \min(\mu f_i^{\mathrm{nor}}, f_i^{\mathrm{tang}}).
\label{eq-contact-rate-12}
\end{equation}
where $\mu$ is the coefficient of friction. To complete the formulation of
the contact algorithm, a  value $\mathbf{n}_i^{rs}$ of the normal at node $i$
of the computational mesh for the contacting bodies $r$ and $s$ is still
needed. As an approximation, the following algorithm is presented to
determine the normal value,

\begin{enumerate}
\item If bodies $r$ and $s$ contact at node $i$, initialize vectors $\mathbf{%
V}_r$ and $\mathbf{V}_s$ as zero.

\item Search within the eight cells(for three-dimensional cases) around the
node. If cell $j$ possesses particles belonging to body $r$ (or $s$),
calculate the difference of coordinates between the node $i$ and the center
of cell $j$, $\mathbf{x}_{i}-\mathbf{x}_{c}^{j}$, then add the difference to
vector $\mathbf{V}_{r}$ (or $\mathbf{V}_{s}$).

\item Calculate the difference between the vectors $\mathbf{V}_r$ and $%
\mathbf{V}_s$, then set the difference as the value of $\mathbf{n}_i^{rs}$.
Finally, unitize $\mathbf{n}_i^{rs}$.
\end{enumerate}

Finally, we summarize the material point method with the new multi-mesh
contact algorithm presented in this paper as follows:

\begin{enumerate}
\item Get the particle mass $m_r$, position $\mathbf{x}_r$,velocity $\mathbf{%
v}_r$, density $\rho_r$, and stress $\mathbf{\sigma}$; form the lumped mass
matrix(Eq. \eqref{eq-pmp-mass-2}) and nodal momentum(Eq. %
\eqref{eq-pmp-mass-velo-2}).

\item Loop over the mesh nodes, if two bodies contact at node $i$, adjust
the nodal momenta of the contacting bodies(Eq. \eqref{eq-contact-rate-4a},
Eq. \eqref{eq-contact-rate-4b}).

\item Calculate the rate of the deformation gradient for each particle,
compute the increment of strain using an appropriate strain measure and
solve constitutive equations to update the stress, $\mathbf{\sigma}_p$.

\item Form the internal force (Eq. \eqref{eq-pmp-fint}). Calculate the
contact force between bodies and form the external force(Eq. %
\eqref{eq-pmp-fext})

\item Solve the momentum equations for the nodal accelerations and get the
velocity in a Lagrangian frame:
\begin{equation}
m_{i} [ \mathbf{v}_i|_{t+\Delta t} - \mathbf{v}_i|_t ]= \Delta t[ (\mathbf{f}%
_i)^{\mathrm{int}} + (\mathbf{f}_i)^{\mathrm{ext}}]  \label{eq-pmp-solve-2}
\end{equation}

\item Update the solution at the material point by mapping the nodal values
using the element shape functions. Positions and velocities are updated
according to
\begin{equation}
\mathbf{x}_p|_{t+\Delta t} = \mathbf{x}_p|_{t} + \Delta t \sum_{i}{\ \mathbf{%
v}_i|_{t+\Delta t} N_i(\mathbf{x}_p) }  \label{eq-pmp-pos}
\end{equation}
and
\begin{equation}
\mathbf{v}_p|_{t+\Delta t} = \mathbf{v}_p|_{t} + \sum_{i}{[ \mathbf{v}%
_i|_{t+\Delta t} - \mathbf{v}_i|_{t} ]N_i(\mathbf{x}_p) }
\label{eq-pmp-velo-update}
\end{equation}

\item Define a new finite element mesh if necessary, and return to step 1 to
begin a new time step.
\end{enumerate}

\section{Numerical Simulation\label{SEC-NUMERICAL}}

Numerical simulations presented in this section are carried out in three
dimension. The first set of simulation involves a cylinder rolling on an
inclined rigid plane and is meant as simple illustration and validation of
the friction algorithms presented by Bardenhagen et al. and by us. The
second set involves the collision of two elastic spheres and is to examine
the efficiency of the multi-mesh contact algorithm proposed in this paper.
The third set involves a copper Taylor bar impacting to a rigid wall. The
last example is to simulate the process of the collision between four
identical spheres. The last two examples examine the conservation of energy
during the collision is checked.

\subsection{Rolling simulation\label{SEC-NUM-ROLLING}}

Fig \ref{Fig1} shows the plane geometry for a computation with a cylinder on
an inclined plane. In this example, the plane inclined at an angel $\theta$
with respect to the horizontal line, while the gravity $\mathbf{g}$ is
vertically downward.
\begin{figure}[tbp]
\centering
\caption{Geometry for simulations of a cylinder on an inclined plane}
\label{Fig1}
\end{figure}

A rigid cylinder on an inclined surface will roll, or slip depending on the
angle of inclination and friction coefficient. Specifically, if $\tan{\theta}%
>3\mu$, the cylinder will roll and slip; Otherwise the cylinder will roll
without slipping, where $\mu$ is the coefficient of friction. For an
initially stationary, rigid cylinder, the $x$-component and the
center-of-mass position as a function of time, $x_{\mathrm{cm}}(t)$, is
given by
\begin{equation}
\begin{array}{l}
x_{\mathrm{cm}}(t) = \left\{
\begin{array}{ll}
x_0 + \frac{1}{2}|\mathbf{g}|t^2(\sin \theta - \mu \cos \theta), & \tan
\theta > 3\mu \hspace{5mm} (\mathrm{slip}), \\
x_0 + \frac{1}{3}|\mathbf{g}|t^2\sin\theta , & \tan \theta \le 3\mu \hspace{%
5mm} (\mathrm{stick}),%
\end{array}
\right.%
\end{array}
\label{eq-example-incline-1}
\end{equation}
In Eq. \eqref{eq-example-incline-1}, $x_0$ is the $x-$component of the
initial center-of-mass position, and $|\mathbf{g}|$ is the magnitude of the
gravitational acceleration.

Simulation is performed with a cylinder that has the radius $R=40\mathrm{mm}$%
, thickness $t=20\mathrm{mm}$, and gravitational acceleration with magnitude
$10\mathrm{m/s^2}$. The computational domain is cubic with side length
length $700\mathrm{mm}, 150\mathrm{mm}$ and $40\mathrm{mm}$, respectively.
The cell size is $10\mathrm{mm}$ so there are only eight and two
computational elements across the diameter and the thickness of the
cylinder, respectively. The simulation involves a elastic, deformable
cylinder with elastic modulus $1.24$MPa, Poisson ratio 0.35 and density $%
8.0\times 10^{-3}\mathrm{g}/\mathrm{mm}^3$. The inclined plane is
discretized as a rigid body with $70$, $15$ and $4$ material points,
respectively, and there is only one material point in one cell.

\begin{figure}[tbp]
\centering
\includegraphics*[width=1.0\textwidth]{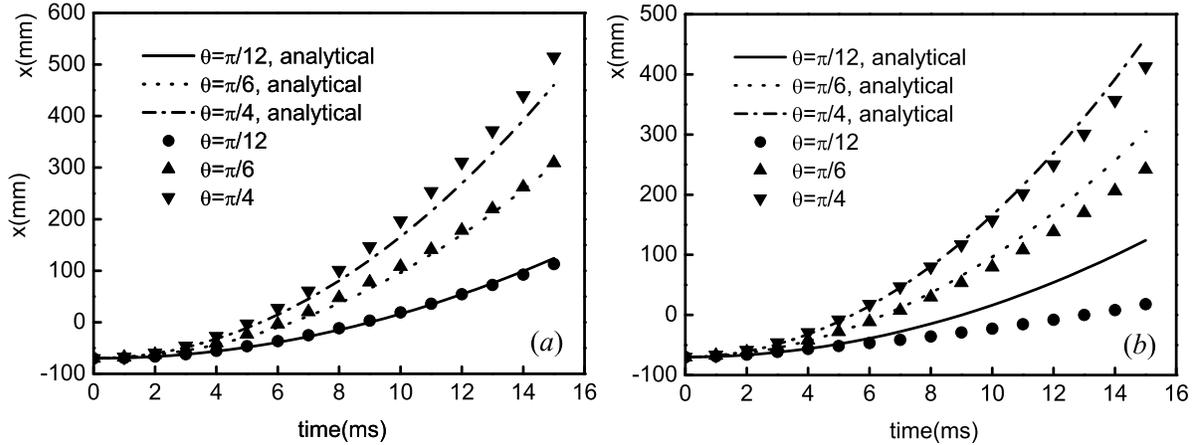}
\caption{Center-of-mass position for deformable cylinder vs time. The
coefficient of friction, $\protect\mu =0.5$, the angles of inclination, $%
\protect\theta =\protect\pi /12$, $\protect\pi /6$ and $\protect\pi /4$,
respectively. The symbols represent simulation results while lines represent
analytical ones. ($a$)our contact algorithm,($b$) contact algorithm by
Bardenhagen.}
\label{Fig2}
\end{figure}

Fig. \ref{Fig2} shows the center-of-mass position of the cylinder as a
function of time for three values of angel of inclination, $\theta =\pi /12$%
, $\theta =\pi /6$ and $\theta =\pi /4$, respectively, and the coefficient
of friction fixed at $\mu =0.5$. The symbols represent simulation results,
and lines represent analytical ones. Fig. \ref{Fig2}($a$) shows the results
of our contact algorithm while Fig. \ref{Fig2}($b$) shows those of contact
algorithm by Bardenhagen. For cases with large inclination angle the results
of both contact algorithms agree well with analytical solutions. But when
the inclination angle is small, results of our contact algorithm are much
better.

\begin{figure}[tbp]
\centering
\includegraphics*[width=1.0\textwidth]{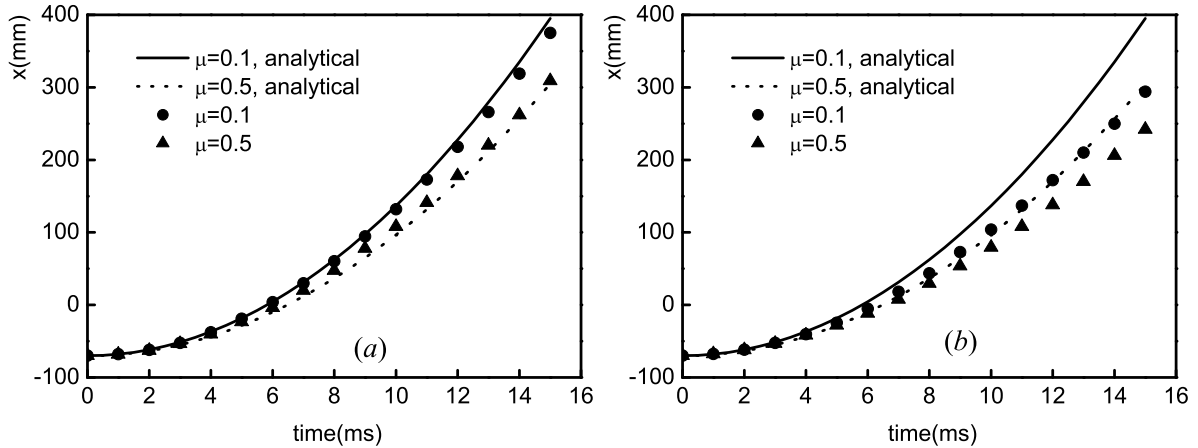}
\caption{Center-of-mass position for deformable cylinder as a function of
time. The angle of inclination, $\protect\theta = \protect\pi/6$, and
coefficient of friction, $\protect\mu = 0.1$, $0.5$, respectively. The
symbols represent simulation results while lines represent analytical ones. (%
$a$) our contact algorithm, ($b$) algorithm by Bardenhagen.}
\label{Fig3}
\end{figure}

In the next test, the value of angle of inclination is fixed at $\theta=\pi/6
$ and the coefficient of friction is varied, $\mu=0.1$ and $\mu = 0.5$. Fig. %
\ref{Fig3} shows the center-of-mass position of the cylinder as a function
of time for each simulation and the corresponding exact solution for a rigid
cylinder. The computed results for the deformable cylinders agree well with
the analytical solutions, and as before, the computed curves obtained with
our contact algorithm are much more closer to the analytical curves than
those by Bardenhagen's algorithm.

\begin{figure}[tbp]
\centering
\includegraphics*[width=1.0\textwidth]{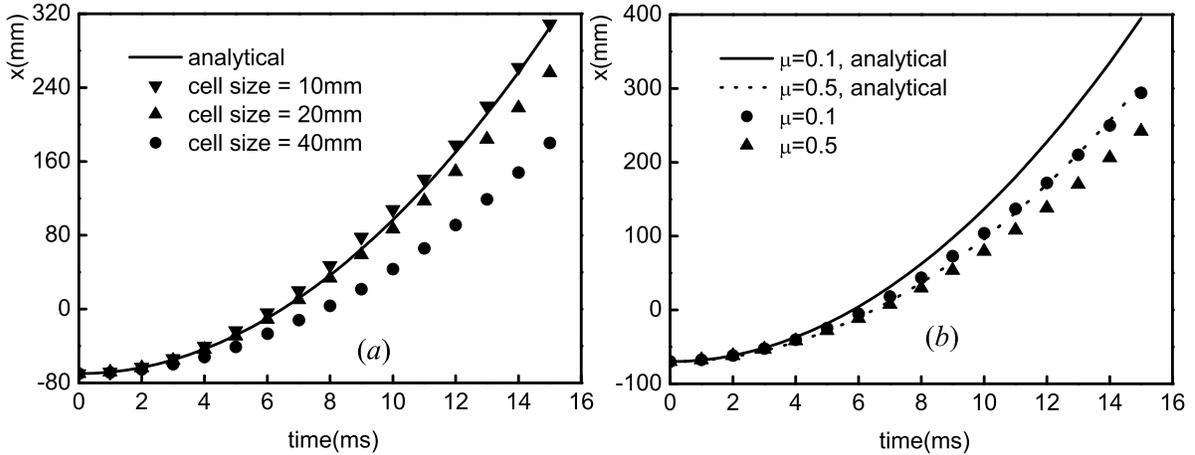}
\caption{Center-of-mass position for deformable cylinder as a function of
time. In the simulation angle of inclination is $\protect\theta =\protect\pi %
/6$, coefficient of friction is $\protect\mu =0.5$ and side lengths of cells
are $40\mathrm{mm}$, $20\mathrm{mm}$ and $10\mathrm{mm}$, respectively. ($a$%
) our algorithm, ($b$) algorithm by Bardenhagen.}
\label{Fig4}
\end{figure}

Fig. \ref{Fig4} shows simulation results for different mesh sizes. In this
test, the angle of inclination is fixed at $\theta =\pi /6$ and the
coefficient of friction is fixed at $\mu =0.5$. The side length of the cubic
computational elements is varied, $40\mathrm{mm}$, $20\mathrm{mm}$ and $10%
\mathrm{mm}$, respectively. Fig. \ref{Fig4}($a$) shows the results of our
contact algorithm while Fig. \ref{Fig4}($b$) shows those of Bardenhagen's.
It is clear that the simulation results converge to the analytical ones with
the decrease of the mesh size. The agreement between results by our contact
algorithm agree better with analytical ones than those by Bardenhagen's
scheme in the later time.

\subsection{Impact of elastic spheres\label{SEC-NUM-SHPERES}}

In this section, the impact of two elastic spheres is simulated to test the
conservation of the energy during the impact with the contact algorithm. The
variables are all dimensionless in this example. The spheres start from the
left side and the center of the computational domain with initial velocities
$(0.1,0,0)$ and $(0,0,0)$, respectively. The computational domain is a cube
whose sides along the $x$, $y$ and $z$ direction are $40$, $20$ and $20$,
respectively, and cubic meshes are used with side length $\Delta x = \Delta
y = \Delta z = 0.5$. Eight material points are used per element. The spheres
have a radius of $4$, Young's modulus of $1000$, a Poisson's ratio of $0.3$
and a density of $1.0$. The distance between the center of first sphere and
that of the second sphere is 14. The simulation is run up to a final time $%
t=80$.

The results from our contact algorithm are shown in Fig.\ref{Fig5}. As a
comparison, the results from the contact algorithm by Bardenhagen are shown
in Fig.\ref{Fig6}. In order to show the moving of spheres clearly, only the
central layer of the 3D configuration is shown. The nonlinear constitution
for large-deformation is used. In these figures, ($a$)show the two spheres
at time $t=0.0$ when they just begin to travel through the grid, ($b$)show
the impact of spheres at time $t=37.8$ and ($c$)show the spheres at time $%
t=75.0$. From Fig.\ref{Fig5} ($c$) we find two spheres separate, and the
left one is almost immobile and the right one moves with nearly the same
kinetic energy as the initial kinetic energy of left one. But in Fig.\ref%
{Fig6} ($c$) the two spheres move forward together, which is unphysical.

\begin{figure}[tbp]
\centering
\caption{Snap-shots of impact of two elastic spheres obtained by our contact
algorithm. From up to down, the corresponding times are $0\mathrm{ms}$, $37.8%
\mathrm{ms}$ and $75.0\mathrm{ms}$, respectively.}
\label{Fig5}
\end{figure}

\begin{figure}[tbp]
\centering
\caption{Snap-shots of impact of two elastic spheres obtained by
Bardenhagen's contact algorithm. From up to down, the corresponding times
are $0\mathrm{ms}$, $37.8\mathrm{ms}$ and $75.0\mathrm{ms}$, respectively.}
\label{Fig6}
\end{figure}

\begin{figure}[tbp]
\centering
\includegraphics*[width=0.6\textwidth]{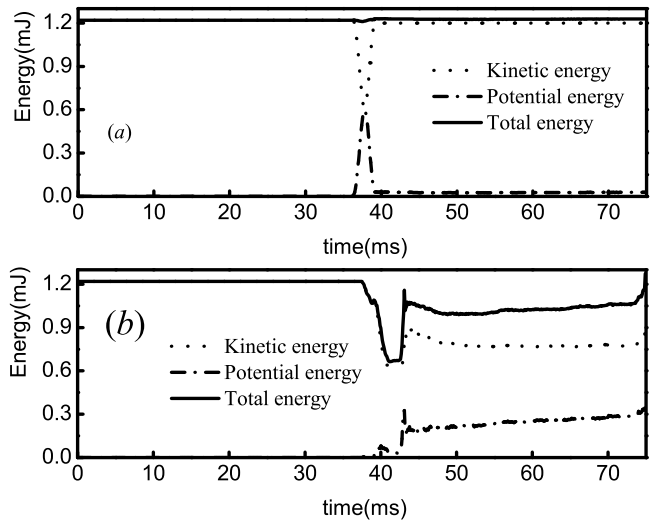}
\caption{Energy evolution of elastic cylinder impact. The results with the
contact algorithms presented in this paper and by Bardenhagen are shown in ($%
a$) and ($b$), respectively. }
\label{Fig7}
\end{figure}

Fig.\ref{Fig7} shows the kinetic, potential and total energies as a function
of time in which ($a$) shows the results of our contact algorithm while ($b$%
) shows those of Bardenhagen's. From Fig. \ref{Fig7}($a$) we find the
kinetic energy decreases during the impact and recovers after the spheres
separate. The potential energy (broken line) begin to accumulate upon impact
of spheres, reaches its maximum value at the point with maximum deformation
during impact, then decreases to a small mount associated with the free
vibration of the spheres after separation. The total energy(solid line) is
approximately constant. In Fig. \ref{Fig7}($b$), the kinetic energy
decreases during the impact but doesn't recover to the original one; the
potential energy begin to increase when the contact begins, then reaches its
a maximum value, and does not decrease. Total energy shown in Fig. \ref{Fig7}%
($b$) is not constant, which shows a strong numerical dissipation during the
course of impact.

\subsection{Impact of a Taylor Bar \label{SEC-NUM-TAYLOR}}

The classical Taylor bar problem is considered. This is a commonly simulated
problem and is often used as a benchmark for transient dynamics computer
codes. A copper bar of radius $R=3.8\mathrm{mm}$ and length $L_{0}=25.4%
\mathrm{mm}$ impacts on a rigid, frictionless wall with an initial
longitudinal velocity of $190\mathrm{m/s}$. The material is modeled as
elastoplastic with Young's modulus $E=117\mathrm{GPa}$, Poisson $\nu =0.35$,
the yield stress is $\sigma _{y}=0.157\mathrm{MPa}$ and linear hardening is
assumed with $H=0.425\mathrm{MPa}$. The material has a density of $\rho
_{0}=8.93\mathrm{g/{cm^{3}}}$. In order to compare the computed results to
those of experiments, we use the following estimation of error given by G.
R. Johnson\cite{Johnson1988}:
\begin{equation}
\bar{\Delta}=\frac{1}{3}\left( \frac{|\Delta L|}{L}+\frac{|\Delta D|}{D}+%
\frac{|\Delta W|}{W}\right)   \label{eq-num-taylor-0}
\end{equation}%
where $L$ and $D$ are the length and diameter of the bottom after the
impact, respectively, as shown in Fig. \ref{Fig8}. $W$ is the diameters of
the layer which is $0.2L_{0}$ to the bottom.
\begin{figure}[tbp]
\centering
\caption{The sketch figure of the impact of a Taylor bar}
\label{Fig8}
\end{figure}

\begin{figure}[tbp]
\centering
\includegraphics*[width=0.6\textwidth]{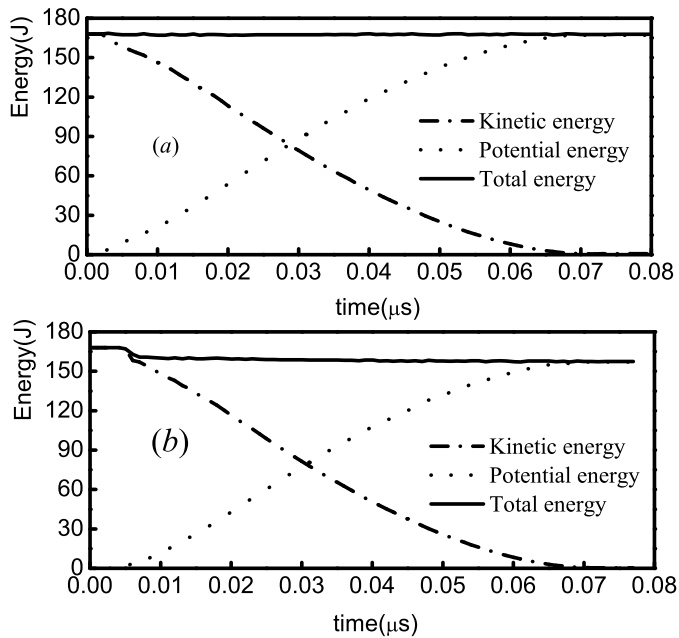}
\caption{Energy evolution during impact of Taylor bar}
\label{Fig9}
\end{figure}

The bar moves within the cubic domain $[-10.4,10,4]\times [-10.4,10.4]\times[%
-2,36]$, meshed by $30 \times 30 \times 50$ elements. For the initial
construction of the bar, there are $8$ material points in every cell, and
for the rigid wall there is only one material point. The unit of coordinate
is millimeter. The terminal time is $80\mathrm{\mu s}$. Fig. \ref{Fig9}
shows the kinetic, potential and total energy as a function of time, where
the contact algorithm presented by us(Fig. \ref{Fig9}($a$)) and by
Badenhagen(Fig. \ref{Fig9}($b$)) are used. In Fig. \ref{Fig9}($a$), the
kinetic energy decreases during the impact and is totally converted to
potential energy at the end. The total energy is constant during the whole
time. In Fig. \ref{Fig9}($b$), the energy is dissipated during the impact.

Table \ref{table-taylor-1} shows the comparison between the computed results
and experimental ones, where MPM1 is MPM with our contact algorithm and MPM2
is MPM with contact algorithm presented by Bardenhagen. The results by MPM1
agree better with experimental ones.
\begin{table}[tbp]
\caption{The comparison between the computed results and experimental ones}
\label{table-taylor-1}{\small \ }
\par
\begin{center}
{\small \
\begin{tabular}{|c|c|c|c|c|}
\hline
& L($\mathrm{mm}$) & D($\mathrm{mm}$) & W($\mathrm{mm}$) & $\bar{\Delta}$ \\
\hline
Experiment & 16.2 & 13.5 & 10.1 & - \\ \hline
MPM1 & 16.15 & 13.21 & 9.63 & 0.071 \\ \hline
MPM2 & 16.25 & 11.96 & 9.42 & 0.184 \\ \hline
\end{tabular}
}
\end{center}
\end{table}

Fig. \ref{Fig10} shows the final particle configuration, colored by contour
values of equivalent plastic strain obtained with MPM1. Fig. \ref{Fig10}($a$%
) shows three-dimensional view while Fig. \ref{Fig10}($b$) shows the center
layer of Fig. \ref{Fig10}($a$) vertically to the rigid wall.
\begin{figure}[tbp]
\centering
\caption{The final particle configuration of the Taylor bar}
\label{Fig10}
\end{figure}

\subsection{Impact of plastic spheres \label{SEC-NUM-P-SHPERES}}

The last example simulates the impact of four identical copper spheres with
the contact algorithm presented in this paper. The material parameters of
spheres are the same as those of last example. The radius of the spheres is $%
10\mathrm{mm}$. Initially, one of the spheres locates at $(0.0, 0.0, 25.0)$
and travels with a velocity $-100\mathrm{m/s}$ parallel to the $z$ axis; The
other three spheres locate at $(10, -5.7735, 0.0)$, $(0.0, 11.547, 0.0)$ and
$(-10.0, -5.7735, 0.0)$ are at rest. The unit of coordinate is millimeter.

\begin{figure}[tbp]
\centering
\includegraphics*[width=0.6\textwidth]{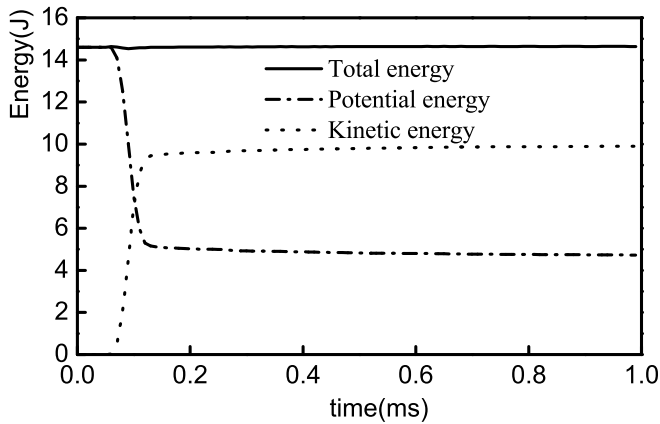}
\caption{Energy evolution during impact of four copper spheres. }
\label{Fig11}
\end{figure}

The spheres moves within the cubic domain $[-50,50]\times [-50,50]\times[%
-50,50]$, meshed by $50 \times 50 \times 50$ elements. Fig. \ref{Fig11}
shows the kinetic, potential and total energy as a function of time. The
kinetic energy decreases during the impact and part of the kinetic energy is
converted to potential energy during the impact. The total energy is a
constant during the whole time.

Fig.\ref{Fig12}($a$)--($c$) show the particle configurations of different
time, colored by contour value of equivalent plastic strain. From blue to
red the plastic strain increases and correspondingly the possible
temperature increment becomes larger. Fig. \ref{Fig12}($a$) shows the
initial particle configuration at $t=0\mathrm{\mu s}$. Fig. \ref{Fig12}($b$)
shows the particle configuration at $t=80\mathrm{\mu s}$ when the upper
sphere just impact to the lower three ones. Fig. \ref{Fig12}($c$) show the
particle configuration at $t=1\mathrm{ms}$ when they separate.

\begin{figure}[tbp]
\centering
\caption{Snapshots of impact of four copper spheres. From black
to white the plastic strain increases and correspondingly the possible
temperature increment becomes larger. ($a$)$t=0\mathrm{\protect\mu }s$; ($b$)%
$t=80 \mathrm{\protect\mu s}$; and ($c$)$t=1 \mathrm{ms}$}
\label{Fig12}
\end{figure}

\section{Conclusion\label{SEC-CON}}

A new multi-mesh contact algorithm for three-dimensional material point
method is presented. The contact algorithm faithfully recovers  the opposite
acting forces between colliding bodies. Collision procedures between regular
bodies and/or rigid bodies can be treated within the same framework.  A
multi-value of momentum and mass is defined on every node to describe the
contact/sliding/separation procedure. Both normal and tangential  velocities
of each particle at the contact surface are calculated in respective
individual mesh. A  Coulomb friction is applied to describe the sliding or
slipping  between the contacting bodies. The efficiency of the contact
algorithm is linearly related to the number of the contacting  bodies
because the overlapped nodes are labeled by sweeping the material particles
of all bodies when the nodal momentum and mass  are formed in every time
step.

Numerical simulation shows that our contact algorithm possesses high
accuracy and low numerical  energy dissipation, which is very
important  for solving collision problems.

\section*{Acknowledgments}

We warmly thank G. Bardenhagen, Haifeng Liu, Song Jiang, Xingping
Liu, Xijun Yu, Zhijun Shen, Yangjun Ying, Guoxi Ni, and Yun Xu for
helpful discussions.

\end{document}